\documentclass[conference]{IEEEtran}
\usepackage{fancyhdr}
\usepackage{graphics}
\usepackage{listings}
\usepackage{xspace}
\usepackage{amsmath}
\usepackage{algorithm}
\usepackage{multirow}
\usepackage{rotating}
\usepackage{hhline}
\usepackage{blindtext}
\usepackage[noend]{algpseudocode}
\usepackage{etoolbox}%
\usepackage{enumitem}
\usepackage{soul}
\usepackage[colorinlistoftodos]{todonotes}
\usepackage{xcolor}
\usepackage[most]{tcolorbox}
\usepackage{setspace}
\usepackage{times}
\usepackage[T1]{fontenc}
\usepackage{amsmath}

\makeatletter
\newcommand*{\rom}[1]{\expandafter\@slowromancap\romannumeral #1@}
\makeatother

\usepackage{xr}
\definecolor{mygreen}{rgb}{0,0.6,0}
\newcommand{\name}{\texttt{FlipTracker}\xspace}
\newcommand{\subheader}[1]{ \textbf{#1}}

\newcommand\note[1]{\todo[inline, size=\small, color=blue!40]{Comment: #1}}

\newcommand{\squeezeup}{\vspace{-3mm}}

\usepackage[normalem]{ulem}

\usepackage{tikz}
\newcommand*\circled[1]{\tikz[baseline=(char.base)]{
            \node[shape=circle,draw,inner sep=2pt] (char) {#1};}}

\ifCLASSINFOpdf
\else
\fi
\begin{document}

\title{
FlipTracker: Understanding Natural Error Resilience in HPC Applications
}

\author{\IEEEauthorblockN{Luanzheng Guo}
\IEEEauthorblockA{EECS, UC Merced \\
lguo4@ucmerced.edu}
\and
\IEEEauthorblockN{Dong Li}
\IEEEauthorblockA{EECS, UC Merced \\
dli35@ucmerced.edu}
\and
\IEEEauthorblockN{Ignacio Laguna}
\IEEEauthorblockA{Lawrence Livermore National Laboratory\\
ilaguna@llnl.gov}
\and
\IEEEauthorblockN{Martin Schulz}
\IEEEauthorblockA{Technical University of Munich\\
schulzm@in.tum.de}
}

\maketitle

\thispagestyle{fancy}
\lhead{}
\rhead{}
\chead{}
\lfoot{\footnotesize{
SC18, November 11-16, 2018, Dallas, Texas, USA
\newline 978-1-5386-8384-2/18/\$31.00 \copyright 2018 IEEE}} \rfoot{}
\cfoot{} \renewcommand{\headrulewidth}{0pt} \renewcommand{\footrulewidth}{0pt}

\pagestyle{plain}


\begin{abstract}
As high-performance computing systems scale in size and computational power,
the danger of silent errors,
i.e., errors that can bypass hardware
detection mechanisms and impact application state, {\color{black} grows} dramatically.
Consequently, applications running on {\color{black} HPC} systems need to
exhibit resilience to such errors. Previous work has found
that, for certain codes, this resilience can come for
free, i.e., some applications are naturally resilient, but
few studies 
have shown the code patterns---combinations or sequences of 
computations---that make an application naturally resilient.
In this paper, we present \name, a framework designed to 
extract these patterns using fine-grained tracking of error propagation and 
resilience properties, and we use it to present a set of computation patterns that 
are responsible for making representative HPC applications 
naturally resilient to errors. 
This not only 
enables a deeper understanding of resilience properties of 
these codes, but also can guide future application designs towards  patterns with natural resilience. 
\end{abstract}


%
\IEEEpeerreviewmaketitle

\section{Introduction}
\label{sec:intro}
Ensuring execution correctness and result integrity in High-Performance 
Computing (HPC) simulations is an urgent need in extreme-scale systems.
As systems scale and the number of system components grow,
the chances of experiencing errors
increases as well~\cite{snir2014addressing}. Although most soft errors---transient 
faults that are induced by electrical noise or external high-energy 
particle strikes---can be detected and corrected by hardware- and 
system-level mechanisms, some errors can escape these mechanisms and 
propagate to the application. These \textit{silent errors} can then 
generate Silent Data Corruption (SDC), impacting scientific results 
without users realizing it.

As the probability of SDC grows, it becomes increasingly 
necessary to develop 
applications that can transparently tolerate, or \textit{mask}, these errors
before they affect the application's numerical output.
Previous work on fault tolerance, which typically focused on individual applications,
demonstrates that a number of applications
have this property and can mask errors as they appear.
Examples of such applications are algebraic multi-grid
solvers (AMG)~\cite{ics12:casas}, Conjugate Gradient (CG) solvers~\cite{Sao:2013},
GMRES iterative solvers~\cite{Elliott:2014},
Monte Carlo simulations~\cite{sc-15:ashraf}, and machine learning algorithms,
such as clustering~\cite{leem2010} and deep-learning neural 
networks~\cite{alippi1995sensitivity, piuri-2001analysis}.

While previous work attributes this natural resilience at a high-level 
to either the probabilistic or iterative nature of the
application, the community still lacks the fundamental understanding on 
the program constructs that result in such natural 
error resilience. 
Fundamentally, we do not have clear answers
to questions, such as:
Are there any common computation patterns (i.e., combinations or 
sequences of computations) that lead to 
natural error resilience?
If so, how can these patterns be found?
How can future application design benefit from 
patterns exhibiting natural resilience?
Finding answers to these questions is critical 
for error 
detection and recovery to avoid overprotecting regions of code 
that are naturally resilient.

In this paper, we characterize application natural resilience
using common HPC programs 
and identify 
six common resilience 
computation patterns. Examples of such patterns are 
\textit{dead corrupted variables}, where sets of corrupted 
temporal variables are not used afterwards, and \textit{repeated additions}, 
a pattern that amortizes the effect of incorrect data values.

To capture and extract these patterns, however, a new method is required.
While some methods exist to inject faults
and statistically quantify their manifestation, such as
%
\textit{random fault injection}~\cite{ics12:casas, europar14:calhoun, 
 bifit+:SC12, sc16:guanpeng, prsdc13:sharma}, and to use \textit{program analysis}~\cite{asplos12:hari, isca14:hari, tdsc11:pattabiraman, prdc05:karthik, hpdc17:Calhoun} to track errors on individual instructions, 
these methods 
miss the fine-grained information on error propagation as well as the context needed to explain, at 
a fine granularity, how errors propagate and consequently how natural resilient computations occur.
%
In other words, these approaches do not provide the needed reasoning about 
how multiple computations work together to make an error disappear 
or to diminish its impact. 



To address the above problems, we design \name, a framework to 
analytically track error propagation and to provide fine-grained understanding of 
the propagation and tolerance of errors in HPC applications, and then apply it
to a series of representative HPC applications to extract the patterns that
provide natural resilience.

Our framework has three key features. \textit{First}, we introduce
an application model that partitions the application into code regions.
Such a model allows us to build a high-level picture 
on how an error propagates across code regions, or is tolerated 
with the combination of multiple code regions.
\textit{Second}, using data dependency analysis, we identify the
input and output variables of each code region, which allows
us to perform isolated fault injections at the entry of code regions to 
study their resilience in an isolated fashion. Further, it allows us to quickly
track how the corrupted values change across code regions as caused by 
their resilience computation patterns.
\textit{Third}, we track how the number of live, yet corrupted locations change within code regions, 
an approach that reveals resilience patterns 
that cannot be easily found by traditional high-level fault propagation
approaches.

We present two use cases to demonstrate how resilience computation patterns
can be used to (1) improve application resilience during programming 
and (2) predict the degree of application resilience.

In summary, the contributions of this paper are
\begin{enumerate}[leftmargin=*]

\item An abstract code structure model that enables us to reason 
about the natural resilience properties of code segments;

\item The design of a framework that enables fine-grained and 
comprehensive analysis of error propagation to capture application natural resilience;

\item An implementation of the framework, \name, using the
LLVM compiler and a study of a set of representative HPC programs
on which \name is demonstrated;

\item An analysis and formal definition 
of six resilience computation patterns
that we discover in these programs;

\item Two use cases that demonstrate the usage of 
resilience computation patterns.




\end{enumerate}

\section{Background}
\label{sec:background}
In this section, we define our fault manifestation
model, as well as the concept of resilience computation patterns.

\subsection{Fault Model}
We consider soft errors, also known as 
transient faults,
that propagate to state visible to the application; by \textit{state}
we mean mainly machine registers and memory. We do not consider errors
that are detected, and possibly corrected in hardware, e.g., by 
hardware-level mechanisms such as memory scrubbing, ECC, or other techniques. 
Furthermore, as most other studies in this area \cite{asplos-15:vilas, 
fengshui_sc_13, micro-14:wilkening,asplos12:hari, isca14:hari, micro-16:ventatagiri}, we 
only consider single bit flip errors since it is generally accepted that multi-bit errors are  
much less likely to occur, even in larger HPC systems~\cite{dsn15:schirmeier}.

\subsubsection{Fault Manifestation Model}
\label{sec:manifestations}
We use fault injection to mimic the effect of real soft errors in the
application (Section~\ref{sec:fault_injection} describes our 
fault injection scheme). We define two classes of executions:
\textit{fault-free} runs, on which no fault is injected, and
\textit{faulty} runs on which a fault is injected. When a fault is injected,
we define three possible fault manifestations:
\begin{itemize}[leftmargin=*]

\item \textbf{Verification Success:} in this case, any of two possible scenarios
occur: (a) the program outcome in a faulty-run
is exactly the same as the outcome in a fault-free run; 
or (b) the program outcome in a faulty-run is slightly different from 
the outcome in a fault-free run, but the program successfully passes
the test in its verification phase, i.e., the application 
output is considered correct to the user.

\item \textbf{Verification Failed:} the program terminates, but the outcome 
does not pass the test in the verification phase. This is
a strong indication of SDC that was not tolerated.

\item \textbf{Crashed:} the injected fault generates a crash or a hang.
\end{itemize}

\subsubsection{Success Rate}
\label{sec:success rate}
Success rate is a metric that quantifies application resilience.
In a fault injection campaign, where $M$ fault injection tests are 
performed (see Section~\ref{sec:fault_injection} for details), 
the success rate is defined as
\begin{equation}
\small
\label{eq:suc_rate}
suc\_rate = \frac{\text{\textit{\#Verification Success}}}{M},
\end{equation}
where \textit{\#Verification Success} is the number of 
Verification Success cases of the campaign.
In this paper, we use the success rate as a metric to quantify  application resilience.



\begin{figure*}
	\begin{center}
		\includegraphics[width=0.95\textwidth]{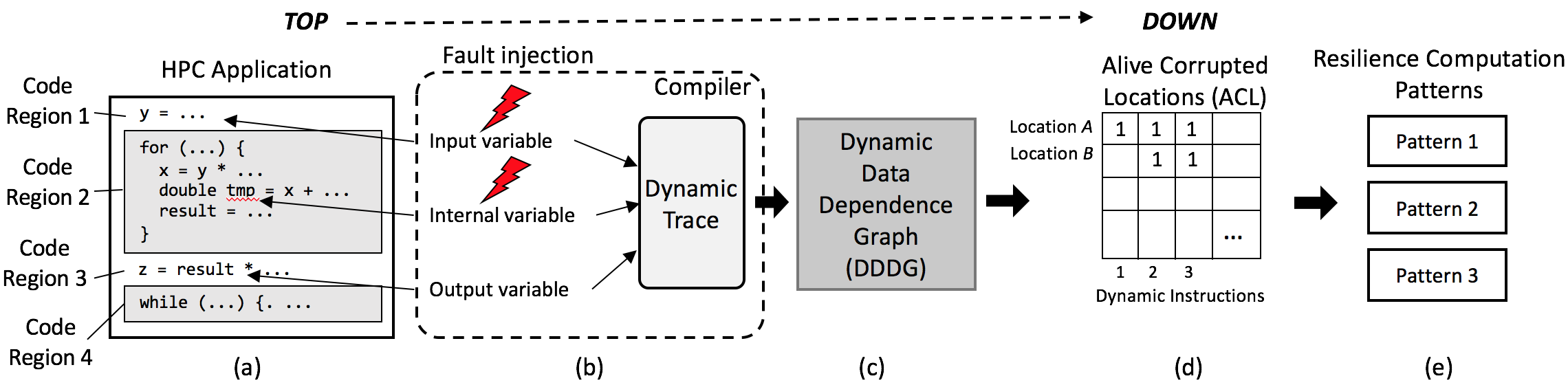} 
		\caption{\small An overview of \name.}
		\label{fig:general_desc}
	\end{center}
    \squeezeup
\end{figure*}

\subsection{Resilience Computation Patterns}

When a fault propagates to application state, it initially corrupts
one (or a few) data locations, i.e., registers and memory locations.
As time passes, instructions that are influenced 
by those corrupted locations can also become corrupted,
causing the total number of corrupted locations in the application
to increase over time. 
Some applications, or code regions of an application, however,
which can tolerate faults, could make the total number of corrupted
locations decrease. 
The above phenomena are depicted in Figure~\ref{fig:lulesh_corruption} in our evaluation section.
If the decrease is sufficient, 
the fault manifests as Verification Success.
Although some applications, or code regions, 
that can tolerate faults do not have such decrease of number of 
corrupted locations, they are characterized with a decrease of 
error magnitude---the relative error of a faulty value with respect to its
correct value.
We say that the above fault tolerant applications or code regions 
are \textit{\textbf{naturally resilient}}.

We define a \textit{\textbf{resilience computation pattern}} as 
a series or a combination of series of computations (or instructions) that 
are responsible for contributing to a decrease of the total 
number of corrupted data locations or a decrease of error magnitude in corrupted data values, and for 
ultimately helping the program tolerate a fault. In this 
paper, we are interested in characterizing the properties of such patterns to answer the following questions:
(a) Why does such a decrease in the number of corrupted locations or error magnitude occur, and (b) What are the patterns that cause this effect?

\section{Design of \name}
\label{sec:design}

In this section, we introduce our method to identify resilience computation patterns.

\name takes as input an HPC program, creates a dynamic execution trace generated
using LLVM instrumentation, and then uses our novel analysis techniques to provide a fine-grained
representation of error propagation and error tolerance. This analysis allows
us to easily identify the resilience computation patterns that may
exist in the program, possibly in different code regions of the program.

Our method is based on a top-level characterization of HPC applications,
which we then use to track error propagation and tolerance at 
a low level. In particular, we model an application as a chain of 
code regions, which work together to produce the final result of the 
application. Each of these code regions can have \textit{input}, 
\textit{output}, and \textit{internal} variables. 
Errors can propagate at any point
in time to any of these variables.

Based on the above application model, we build a dynamic data dependency 
graph (DDDG) from an instruction trace collected at runtime that allows us 
to check the value variation of corrupted 
variables across code region instances (i.e., the top level). Using the DDDG, 
we then build a table, which we call the \textit{alive corrupted 
locations} (ACL) table, that keeps track of the corrupted locations 
for each dynamic instruction. This table allows us to 
examine the variation 
of the number of alive, corrupted variables to 
identify fault tolerance at the instruction level (i.e., the bottom level).
In the next sections we give more details of each of these steps
(see Figure~\ref{fig:general_desc}).

\subsection{Application Code Region Model} 
We characterize HPC applications as sets of iterative structures 
or loops. In an HPC application, a main computation loop usually 
dominates the application execution time. Within this main loop, 
there are a number of inner loops that are typically used to 
update large data objects (e.g., a mesh structure in computational 
fluid dynamics), and iterative computations are performed to compute
properties of these objects, such as energy of particles.
Figure ~\ref{fig:hpc_app} shows an example of such loop 
program abstractions corresponding to CG~\cite{nas-npb}.

\begin{figure}
\begin{lstlisting}[xleftmargin=.02\textwidth, 
xrightmargin=.02\textwidth]
static void conj_grad() { //called from the main loop 
    ...
    for () { //a first-level inner loop
    	for () { //a second-level inner loop
            for () {...} //a third-level inner loop
        }
    }
    for () {...} //a first-level inner loop
}
\end{lstlisting}
\caption{An example HPC 
application (CG) with iterative structures.}
\squeezeup
\label{fig:hpc_app}
\end{figure}

\subheader{Code Regions.}
Since HPC applications are typically composed of combinations
of loops, we model an application
as a chain of \textit{code regions} delineated by 
loop structures (Step (a) in  Figure~\ref{fig:general_desc}).
A code region can be either a loop or any block of code between
two neighboring loops.
An application can have multi-level nested loops. We allow 
the user to decide at which loop level, 
code regions are defined.
Note that code regions defined 
at different loop levels \textit{only affect} the analysis time (not the analysis correctness) to identify resilient code regions and patterns. 
Code regions defined at the level of innermost loop tend to 
be small and easy for fine-grained instruction level analysis.
However, we can have many of such small code regions, which 
increases our exploration space.
On the other hand,
code regions defined at the level of outermost loop tend to be
large and we have a smaller exploration space of code regions,
but it would be time-consuming for fine-grained instruction 
level analysis.
In our work, we define each of the first-level inner loops as 
a code region. 


\subheader{Code Region Variables.}
Given a code region, we classify the variables within the code 
region as \textit{input} variables, \textit{output} variables, 
and \textit{internal} variables. Input variables are those that are 
declared outside of the code region and referenced in the code region.
Output variables are those that are written in the code region 
and read after the code region. Other variables that the code region
writes to or reads from are internal variables. A code region can 
have many dynamic instances, each of which corresponds to 
one invocation of the code region at runtime. The values of input,
output, and internal variables can vary across multiple instances
of a code region.

\subheader{Rationale Behind the Model.}
Our loop-based model follows the natural way in which HPC programs are
coded and analyzed; HPC programs are composed of a handful of 
high-level loops
where the program spends most of its time.
Our loop-based model also enables a divide-and-conquer 
approach, where we can identify application subcomponents 
that may or may not have resilience patterns. For example, 
in the error propagation analysis, if the
input variables of a code region are not corrupted, one can infer that the region
is not impacted by an error
and we can skip propagation analysis on it.


\begin{figure*}[ht]
	\begin{center}
		\includegraphics[width=0.9\textwidth]{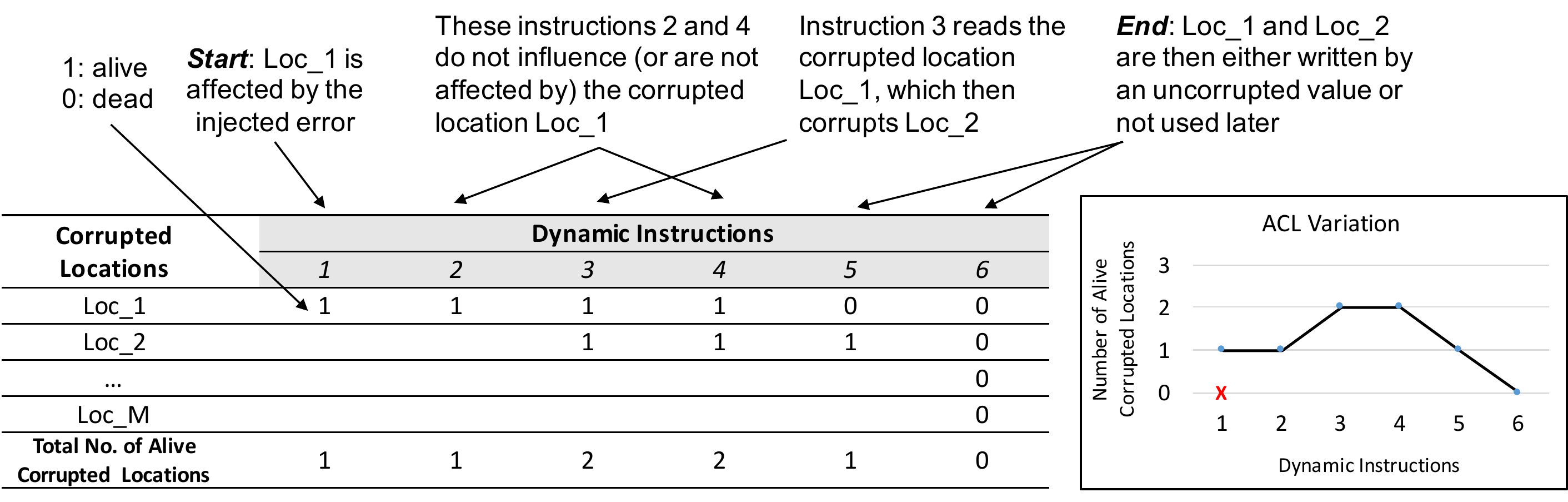} 
		\caption{An example of the ACL table.}
		\label{fig:matrix}
	\end{center}
    \squeezeup
\end{figure*}

\subsection{Tracing Code Region Data}

The DDDG allows us to identify input, output,
and internal variables 
of a code region.
We construct a DDDG for each code region
from a dynamic instruction trace of the application using 
an algorithm inspired by the construction of a program dependence graph~\cite{ferrante1987},
except that our graph is dynamic rather than static:
vertices are the values of 
variables obtained from registers or memory; edges are 
operations transforming input values into output values of variables. 
Using the DDDG as a code region representation, we identify the input
and output variables of the code region: root nodes represent inputs
and leaf nodes represent outputs. Other nodes are internals.

Within the corresponding DDDG of each code region, we inject an error into either the
input, output, or internal variables (Steps (b)--(c) in Figure~\ref{fig:general_desc}).
A DDDG allows us to compare data propagations in regions
with and without fault occurrence, which allows us to
detect control flow divergence by comparing operations.
Further, the values of variables are embedded in the DDDG, 
which helps us to track how specific variables change 
their values across operations; such value change  
reveals whether, how, and where fault tolerance occurs.
%

\subsection{Analyzing Corrupted Variables}
We 
identify variables that, once corrupted, return to their
non-corrupted state and in which dynamic instruction. This
is key in identifying resilience computation patterns since
we need to identify the point in time where the error is tolerated
and its location in the code region 
(Step (d) in Figure~\ref{fig:general_desc}).

Using the DDDG, our analysis of corrupted variables 
gives us a low-level representation in terms of 
\textit{instructions} of how data propagates in the 
code region. Since program abstractions, such as 
variables, are not explicitly represented at 
this level, we need a different way of tracking 
variable values. We introduce a method that 
tracks \textit{alive corrupted locations}, discussed 
as follows. In the following discussion, since a 
variable value can be either in a register location 
or in a memory location, we use the term \textit{location} to cover both options.

\subheader{Alive Corrupted Locations.}
Traversing through the collected instruction trace, we use the DDDG to 
build and dynamically update a table of the \textit{alive corrupted 
locations}, or ACL. Generally speaking, the ACL table stores the number of 
alive, corrupted locations after each dynamic instruction. We call 
a location ``alive'' 
if the value in that location
will be referenced 
again in the remainder of the computation.

Each row of the table shows whether a specific location is alive 
or not after each dynamic instruction, as instructions are encountered in the trace.
Each column of the table shows, for a specific corrupted location, 
whether it is alive or not \textit{after} a dynamic instruction. Based on 
the column information, we can determine the total number of alive, corrupted 
locations after each traced instruction.

Figure~\ref{fig:matrix} gives an example of the ACL table. Each table 
element has a value of 1 or 0, which indicates whether a 
corrupted location after a specific dynamic instruction is alive or not. 
We use the first row as an example to explain the table.  
The location $Loc\_1$ is corrupted by a fault after instruction $1$.
$Loc\_1$ then becomes an alive, corrupted location. Next, $Loc\_1$
remains alive until instruction $5$ where the location is updated and 
the fault in the location is overwritten by a clean value.
The number of alive, corrupted locations are counted after each dynamic
instruction, shown in the last row of the table.


\subsection{Identifying Resilience Patterns from Code Regions}
\label{sec:identifying_code_regions}
As we traverse the instruction trace, the DDDG and ACL table contain the necessary information to 
detect resilient code regions. 
Resilience patterns are extracted from them. 

When the DDDG is used to identify resilient code regions,
we compare
the values of input and output locations in a DDDG between faulty
and fault-free runs. An input location can be corrupted 
\textit{directly}---an error was directly injected into the 
location---or \textit{indirectly}---an error was injected in a 
previous code region, but the error propagates to the input location of the
code region in question. Given a code region, there are two 
possible cases when fault tolerance occurs:

\begin{itemize}[leftmargin=*]
\item 
\textbf{Case 1:} the value of any input location in the code region's
DDDG in a faulty run is incorrect (with respect to the DDDG from a matching
fault-free run), i.e., there is at least one corrupted input location;
however, the values of all output locations are correct. 

\item
\textbf{Case 2:} at least one of the input locations and one of the
output locations in a faulty run are incorrect (with respect to the 
DDDG from a matching fault-free run), but the 
error magnitude in at least 
one corrupted input or output location becomes smaller after 
the code region instance. The error magnitude is defined as 
\begin{equation}
\label{eq:error}
error\_magnitude = \frac{|value_{correct} - value_{incorrect}|}{|value_{correct}|}.
\end{equation}
\end{itemize}

In Case 1, it is reasonable to infer that the code region in question
has natural fault tolerance---the corruption of the input location 
is directly masked within the code region, and does not impact the 
output correctness. 

In Case 2, the error still exists, i.e., there is some amount of 
error in the code region locations; however, the impact of the error, 
measured by its magnitude in the input or output locations, becomes 
smaller, as a function of the code region.
This means that 
the target code region may result in an application outcome
that is numerically different from that of the fault-free executions. 
However, when such a different outcome passes the application verification 
and is acceptable as a valid result, 
we say that Case 2 has fault tolerance. 

When the ACL is used to identify resilient code regions, the algorithm to detect resilience patterns given an ACL is as follows. 
We identify first if in any column, 
an alive corrupted location 
becomes dead 
for a given instruction $i$, where $i < N$ and $N$ is the last instruction before the application outputs 
its result. If this occurs, we mark $i$ as a potential member of 
resilience computation patterns. In Figure~\ref{fig:matrix}, the instruction 5 consuming the location $Loc\_1$ is a potential member of resilience computation patterns.
Once all of such instructions are found, we identify
their source code locations (file and line of code) and provide
them to the user for further analysis.
\section{Implementation}
\label{sec:impl}

We implement \name as a two-step process: first we use a parallel tracer 
built on top of LLVM to extract the instruction traces, and then use 
these traces to dynamically generate and update the DDDGs and the matching 
ACL tables. We do this for both fault-free runs as well as faulty runs. 


\subsection{Parallel Tracing}

\name uses an LLVM instrumentation tool, LLVM-Tracer~\cite{ispass-13:shao},
to generate a dynamic instruction trace. In this trace we store
metadata for each instruction, such as the the instruction type, 
names of registers, and operand values.
In our case, \textit{instructions} 
refer to LLVM instructions, which are generated at the intermediate 
representation (IR) of the program and instrumented by LLVM-Tracer.
This approach does not support MPI programs out-of-the-box, which we need
to support our HPC workloads.
Thus we extend LLVM-Tracer to instrument 
Message Passing Interface (MPI) programs, so that traces are saved 
into a file for each MPI process.

Since trace generation is a per-process task, no synchronization is required
to generate and save per-process traces into different files. Note also
that, in our study, LLVM-Tracer only instruments program
instructions---instructions from the MPI runtime are not instrumented as 
we expect that most errors arise from application computations. 
This however, is not a limitation per se---our approach can easily be directed
to also instrument instructions in any parallel runtime. 
Furthermore, our current implementation can identify errors 
that propagate through MPI communications and then happen in computation, 
even though we do not instrument MPI runtime.

\subheader{Trace Splitting.}
Traces for an HPC program can be quite large for processing. 
Although there is a number of approaches that handle the
problem of large traces (e.g., trace compression~\cite{date07:janapsatya, 
jpdc09:noeth}), we take a simple
approach that splits a trace into smaller pieces. 
Each of small pieces 
corresponds to an instance of a code region, which reduces the scope
for each analysis and further 
allows us to parallelize the analysis. 

\subsection{DDDG Generation and Usage}
Once the trace is generated, \name
takes the dynamic trace as input,
and generates a DDDG by examining the data dependency of the 
operands in each operation. Our technique is based on the 
work of Holewinski et al.~\cite{pldi12:Holewinski},
who proposed a methodology to generate DDDG from a 
dynamic trace.
The generated DDDG is then used to identify the input, internal, 
and output locations for the code region instance 
using Graphviz ~\cite{graphviz}. The DDDG is also used to determine 
corrupted locations by dynamically building the ACL table. 

\textbf{ACL Table Generation.} 
The algorithm to generate an ACL table is motivated by 
dynamic taint analysis in the security
research~\cite{ndss05:newsome,ppopp10:farhana,ccs13:zeng},
which focuses on computations affected by contaminated sources.
The difference between taint analysis and our approach
is that we exclude tainted locations that are never used 
as well as those that are overwritten by an uncorrupted value from 
the untainted location set. In other words, we only consider alive 
corrupted locations in application execution. We use a DDDG 
to acquire the dynamic data dependence to track the error 
propagation, and, simultaneously, we count the number of 
alive corrupted locations after each dynamic instruction 
in the input trace.

\subsection{Fault Injection and Statistical Significance}
\label{sec:fault_injection}
We implement a fault injection framework based on 
FlipIt~\cite{europar14:calhoun},
which allows us to inject a bit flip in the user-specified population of instructions and operands.
Injections are performed randomly
into input and internal locations of code region instances. 
Our fault injection uses a uniformly distributed fault model, similar to~\cite{sc17:giorgis, date09:leveugle}.
Given an input or output location for a code region instance,
we calculate the number of fault injection sites by analyzing the 
dynamic LLVM instruction trace. 
Then, we follow the statistical 
approach in~\cite{date09:leveugle} to calculate the number of fault 
injection tests for a target at 95\% confidence level and 3\% margin of error.

\section{Evaluation}
\label{sec:eval}

\begin{table*}[htbp]
\centering
\small
\caption{Resilience computation patterns in code regions of the HPC programs. DCL, RA, DO represent dead corrupted locations, repeated additions and data overwriting,  respectively.
}
\newcommand{\tabincell}[2]{
\centering
\begin{tabular}
{@{}#1@{}}#2\end{tabular}}
\renewcommand{\arraystretch}{0.85}
\scalebox{0.8}{
\centering
\begin{tabular}{| p{1.5cm} | p{1.0cm} | p{1.4cm} | p{1.8cm} | p{1.0cm} | p{1.0cm} | p{1.0cm} | p{1.0cm} | p{1.0cm} | p{1.0cm} | p{1.0cm} |}
\hline
\textbf{Program} & \textbf{Code region} & \textbf{Line No.} & \textbf{\#instr in an iteration} & \textbf{Pattern\ Found?} &\textbf{DCL} & \textbf{RA} & \textbf{CS} & \textbf{Shifting} & \textbf{Trunc} & \textbf{DO}  \\
\hline
\hline
\multirow{12}* \textbf{$CG$}
& $cg\_a$ & 434-439 & 21017 & NO &  &  &  &  &  &  \\
\cline{2-11}
& $cg\_b$ & 440-453 & 14002 & YES &   & $\surd$ &  &  &  & $\surd$ \\
\cline{2-11}
& $cg\_c$ & 454-460 & 31755757 & YES &  &  & $\surd$ &  &  & $\surd$ \\
\cline{2-11}
& $cg\_d$ & 461-574 & 1196022 & NO &  &  &  &  &  &  \\
\cline{2-11}
& $cg\_e$ & 575-584 & 18202 & NO &  &  &  &  &  &  \\
\hline
\multirow{4}*\textbf{$MG$}
& $mg\_a$ & 425-429 & 606145 & YES &  &  & $\surd$ &  &  & $\surd$ \\
\cline{2-11}
& $mg\_b$ & 430-437 & 719 & YES & $\surd$ & $\surd$ &  &  &  &  \\
\cline{2-11}
& $mg\_c$ & 438-456 & 1019509 & YES & $\surd$ &  &  &  &  & $\surd$ \\
\cline{2-11}
& $mg\_d$ & 457-462 & 3313305 & YES & $\surd$ &  & $\surd$ &  &  & $\surd$ \\
\hline
\multirow{4}*\textbf{$KMEANS$}
& $k\_a$ & 131-142 & 1647 & NO & & & & & & \\
\cline{2-11}
& $k\_b$ & 144-153 & 62 & NO & & & & & & \\
\cline{2-11}
& $k\_c$ & 156-187 & 2185944 & YES &  &  & $\surd$ &  &  & $\surd$ \\
\cline{2-11}
& $k\_d$ & 190-194 & 36 & YES & $\surd$ &  &  &  &  & $\surd$ \\
\hline
\multirow{3}*\textbf{$IS$}
& $is\_a$ & 435-472 & 792630 & NO & & & & & & \\
\cline{2-11}
& $is\_b$ & 473-478 & 983040 & YES &  &  &  & $\surd$ & $\surd$ &  \\
\cline{2-11}
& $is\_c$ & 500-638 &  741367 & YES & $\surd$ &  &  &  &  & $\surd$ \\
\hline
\multirow{1}*\textbf{$LULESH$}
& $l\_a$ & 2652-2693 & 297376 & YES & $\surd$ &  &   &  & $\surd$ & $\surd$ \\
\hline
\end{tabular}
}
\label{tab:ft_code_regions}
\squeezeup
\end{table*}

We apply \name to representative HPC programs to
study their resilience properties and ultimately to extract
naturally resilient patterns that other programs can use.

\subsection{Experimental Setup}
\label{sec:workloads}
We use ten representative HPC programs in our experiments, including 
eight HPC benchmarks (CG, MG, IS, LU, BT, SP, DC, and FT from the NAS 
Parallel Benchmarks in C ~\cite{nas-npb,iiswc:2011} with input Class S), an HPC proxy application (LULESH~\cite{LULESH2:changes} with input ``-s 3"), 
and a benchmark
from the machine learning domain (KMEANS from the Rodinia benchmark suite~\cite{Che:IISWC09} with input ``100.txt").
\subheader{Trace Partitioning and Code Region Selection.}
HPC programs can have several static loop structures, and depending
on program input, each static loop can generate several dynamic instances.
To keep the number of loop instances manageable for analysis,
we focus on high-level loop structures. 
Particularly, we define a code region as a section of the program that is
either (a) a first-level inner loop (if there is any inner loop), or
(b) a code block between two neighbor inner loops.

We list the code regions that we analyzed and
their corresponding line numbers and 
the number of instructions within one iteration of the main loop in 
Table~\ref{tab:ft_code_regions}.

\subsection{Parallel Tracing Overhead}
\begin{figure}[h!]
	\begin{center}
		\includegraphics[width=2.8in]{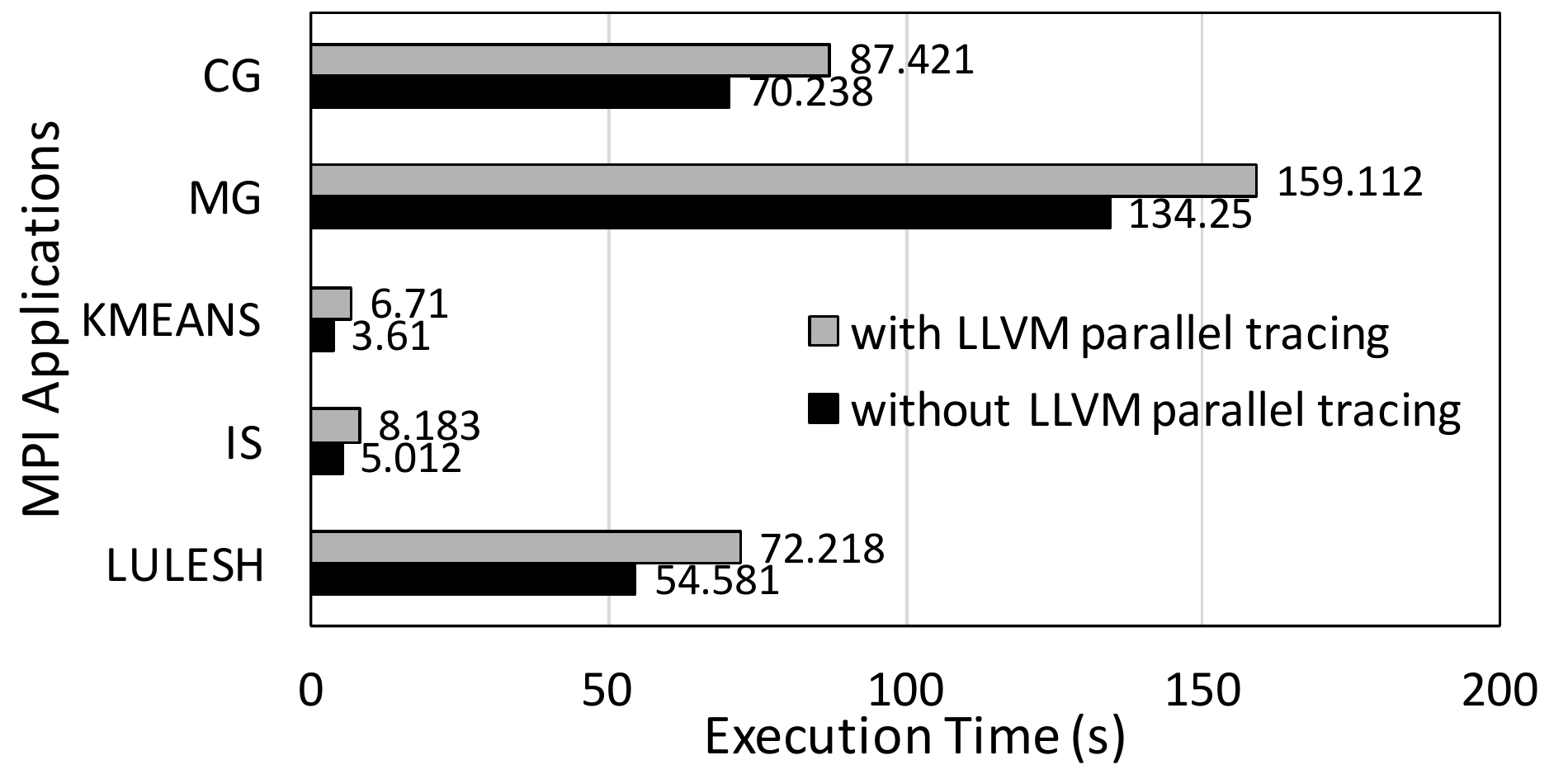} 
		\caption{\small LLVM parallel tracing performance (64 processes on 8 nodes)}
		\label{fig:mpi_overhead}
	\end{center}
    \squeezeup
\end{figure}

We measure the overhead of trace gathering for MPI programs to study
the feasibility of our approach. 
Figure~\ref{fig:mpi_overhead} shows that our approach incurs modest overhead:
45\% on average
when using 64 processes on 8 nodes, comparing to an uninstrumented baseline. 
It is therefore feasible to gather traces at small/medium scales. For large scales, 
one can selectively collect traces for individual functions or 
use techniques such as~\cite{sc06:chung}. We leave the challenge of efficiently
gathering traces at very large scale for future work.


%

Since the resilience computation patterns that we are
interested in occur in the computation code regions
of the program (not in the communication part), 
we focus on the single process 
where the fault is injected.

\subheader{Nondeterminism.}
MPI nondeterminism can bring difficulty to match 
code regions between faulty and fault-free runs.
While in many MPI programs, nondeterminism can be controlled
by eliminating application sources of nondeterminism,
such as calls to \texttt{rand()} and/or \texttt{time()},
in other programs this is difficult because of nondeterminism
introduced by MPI point-to-point communication patterns.
To address these applications, we rely on record-and-replay
tools~\cite{sc15:sato,xue2009mpiwiz}, on which a fault-free 
run is recorded and it is then
replayed in all subsequent faulty executions.




\subsection{Code Region Fault Injection Results}
\label{sec:code-fi-results}
We inject faults in input or internal 
locations of code regions and
measure success rate.
We perform experiments in two dimensions:
(a) across code regions in a given iteration (See ``per-code-region'' results);
(b) in a given code region across all iterations (See ``per-iteration'' results).

\begin{figure*}[th!]
	\begin{center}
		\includegraphics[height=0.117\textheight]{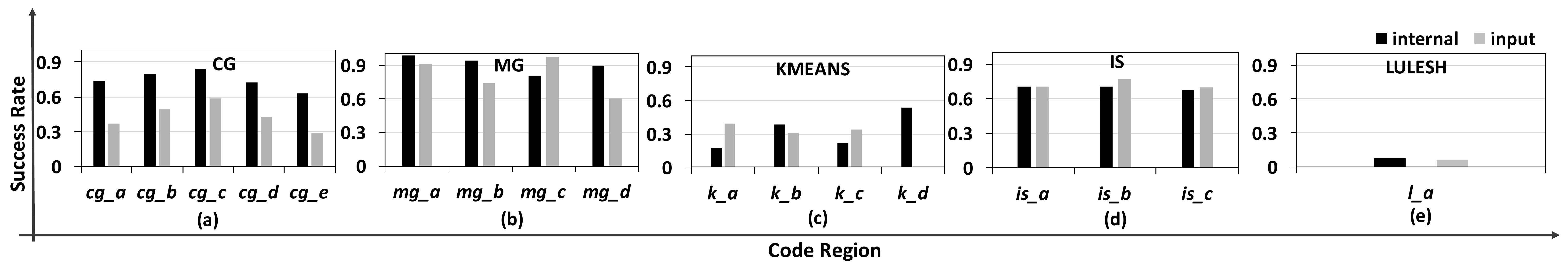} 
		\caption{Fault injection results for code region instances at iteration 0.}
		\label{fig:code_region_variation_in_one_iter}
	\end{center}
    \squeezeup
\end{figure*}

\subheader{Per-Code-Region Results.}
Since different code regions could have different numbers of instances,
to be consistent, we perform the analysis on 
the first instance of each code region, i.e.,
in the iteration $0$ of the main loop
(see Figure~\ref{fig:code_region_variation_in_one_iter}).

\begin{figure*}[th!]
	\begin{center}
		\includegraphics[height=0.117\textheight]{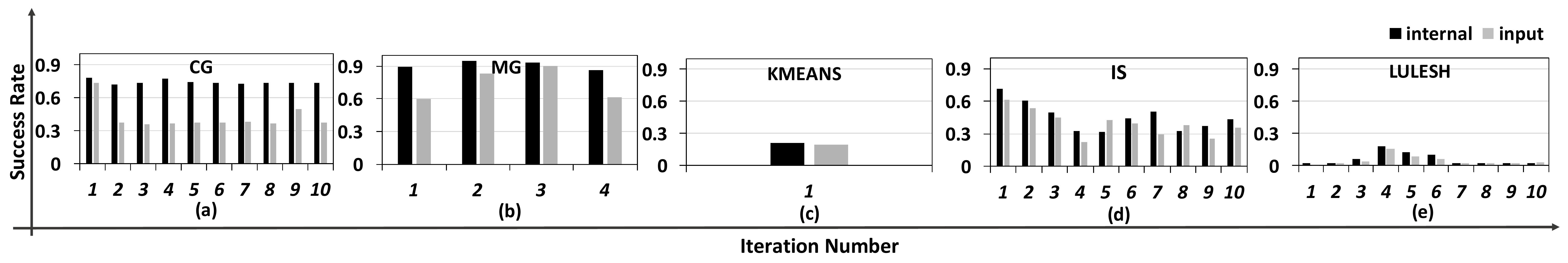} 
		\caption{Fault injection results for individual iterations of the main loop.}
		\label{fig:region_fi}
	\end{center}
        \squeezeup
\end{figure*}

In KMEANS we find that, for faults on internal locations 
the code region k\_d is more resilient than others because
many memory \textit{free} operations free temporal corrupted locations,
while for faults on input locations, many segmentation faults
cause almost zero success rate.
We find a relatively high success rate in MG---we find
cases of repeated addition and dead corrupted location patterns
that account for the fault tolerance (Section~\ref{sec:summary_patterns}
explains these patterns in details).
In IS we find that a bit-shift operation
that occurs on input locations masks faults in the is\_b code region,
which increases its success rate.
In CG, we find two code regions ($b$ and $c$) that have higher
success rates than others because the error magnitudes in variables 
(particularly \texttt{p[]}) become smaller due to a 
computation pattern that repeatedly adds values.
In LULESH, there is only one code region---faults frequently 
cause application crashes, which explains the low 
success rate.

\subheader{Per-Iteration Results.}
We focus on a single code region and
examine its fault tolerance on several loop
iterations. In particular, we treat the main loop of each program
as a single code region and each iteration of the main loop as 
one instance of the code region.
Figure~\ref{fig:region_fi} shows the results.
%
We find that the success rates of different
iterations can be similar. MG (internal locations) and 
CG exemplify this conclusion. The success rates over multiple iterations 
can also be very different, e.g., in IS and LULESH. After examining 
the DDDGs, we find that control flow differences between the 
iterations of the main loop are the main reason accounting for this difference.

\section{Resilience Computation Patterns}
\label{sec:summary_patterns}


We present
a formal description of the resilience computation patterns. 
Table~\ref{tab:ft_code_regions} summarizes them in applications.

\begin{tcolorbox}[breakable,enhanced,left=1mm,right=1mm,top=0.75mm,bottom=0.75mm]
\small
\textbf{Pattern 1: Dead Corrupted Locations (DCL)} \\
In this pattern, the values of several corrupted input locations
are \textit{aggregated} into fewer output locations,
with aggregations being a combination of multiple operations
(e.g., additions and multiplications). While the errors in 
the corrupted input locations can propagate to one (or a few)
locations, many of these corrupted input locations are not
used anymore (they become dead locations) and
the total number of corrupted locations decreases.
\end{tcolorbox}

\begin{figure}[th!]
	\begin{center}
		\includegraphics[width=3.2in]
        {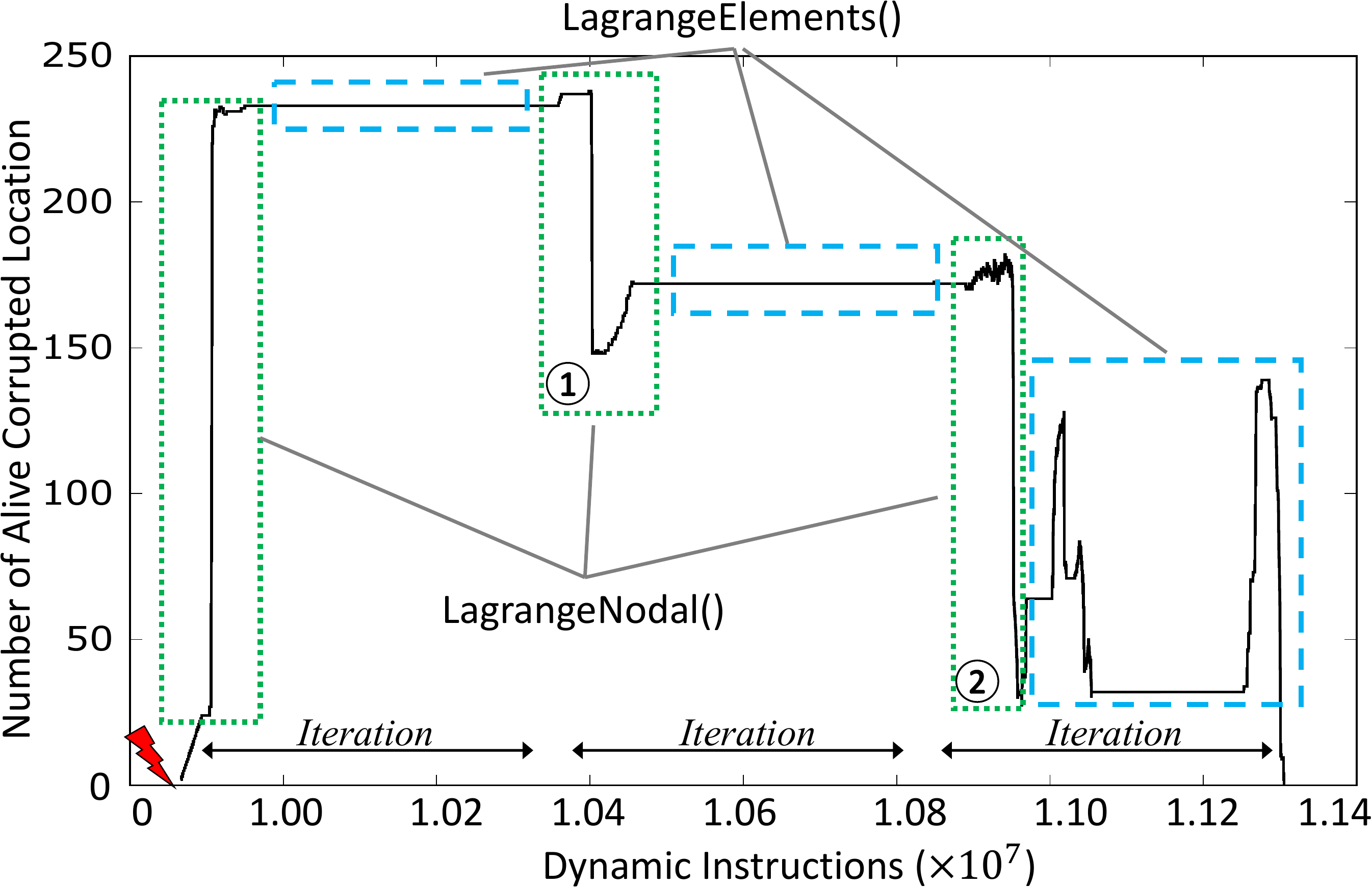} 
		\caption{A real case of ACL table. 
        It shows the number of ACL-s in LULESH after a fault is injected into  
        the last third iteration of the main loop.}
		\label{fig:lulesh_corruption}
	\end{center}
    \squeezeup
\end{figure}

{We frequently 
find Pattern 1 in LULESH.
Figure~\ref{fig:dead_code} shows the code excerpt extracted from LULESH}
that accounts for the decrease of the number of alive corrupted 
locations within the routine \textit{LagrangeNodal} (see \circled{1} 
and \circled{2} in Figure~\ref{fig:lulesh_corruption}). 
The array $hourgram[][]$ is a temporal corrupted location that 
is dead after the sample code snippet. The error has propagated to 
its elements before the example code. Although the error propagates
from $hourgram$ to temporal variables $hxx[]$, which are then aggregated
into $hgfz[]$, the number of alive, corrupted variables decreases
since the corrupted elements of $hourgram[][]$ become dead after this code. We also find this pattern in the MG code.

\begin{figure}
\begin{lstlisting}[caption={}, xleftmargin=.04\textwidth, xrightmargin=.01\textwidth]
for(Index_t i = 0; i < 4; i++){
  hxx[i] = hourgam[0][i]*xd[0]+hourgam[1][i]*xd[1]+
  		   hourgam[2][i]*xd[2]+hourgam[3][i]*xd[3]+
           hourgam[4][i]*xd[4]+hourgam[5][i]*xd[5]+
           hourgam[6][i]*xd[6]+hourgam[7][i]*xd[7];
}
...
for(Index_t i = 0; i < 8; i++) {
      hgfz[i] = coefficient*
                (hourgam[i][0]*hxx[0]+hourgam[i][1]*hxx[1] +
                 hourgam[i][2]*hxx[2]+hourgam[i][3]*hxx[3]);
}
\end{lstlisting}
 \caption{\small Example of the Dead Corrupted Locations in LULESH}
 \label{fig:dead_code}
 \squeezeup
\end{figure}

\begin{figure}
\begin{lstlisting}[label={lst:repeatedly_adding_pattern}, caption={}, xleftmargin=.04\textwidth, xrightmargin=.01\textwidth]
for (i3 = 1; i3 < n3-1; i3++) {
    for (i2 = 1; i2 < n2-1; i2++) {
      ...
      for (i1 = 1; i1 < n1-1; i1++) {
        u[i3][i2][i1] = u[i3][i2][i1]
             +c[0]*r[i3][i2][i1]
             +c[1]*(r[i3][i2][i1-1]+r[i3][i2][i1+1]
                               +r1[i1])
             +c[2]*(r2[i1]+r1[i1-1]+r1[i1+1]);
} } }
\end{lstlisting}
 \caption{\small Example of the Repeated Additions pattern in MG}
 \label{fig:repeated_additions}
 \squeezeup
\end{figure}

\begin{tcolorbox}[breakable,enhanced,left=1mm,right=1mm,top=0.75mm,bottom=0.75mm]
\small
\textbf{Pattern 2: Repeated Additions} \\
In this pattern, the value of a corrupted location is repeatedly 
added by other correct values. Those correct values 
amortize the effect of the incorrect value. This pattern does not 
necessarily cause a decrease of alive, corrupted locations 
(as in Pattern 1), but over time the corrupted value
approaches the correct value such that the application
execution can be successful.
\end{tcolorbox}

We observe Pattern 2 in the iterative solvers MG and CG. 
Figure~\ref{fig:repeated_additions} shows a code excerpt covering  
this pattern in MG. Here, we inject a fault in an element 
of the array $u$ and then the array element $u[i3][i2][i1]$ 
is added with new data values (Lines 6-9). This code 
is repeatedly executed in the main computation routine ($mg3P$).
As a result, the array element $u[i3][i2][i1]$ is repeatedly 
added along with new data values.

\begin{table}
\centering
\begin{center}
\caption{The repeated additions pattern takes effect in MG}
\scriptsize
\label{tab:mg_repeatedly_adding}
\begin{tabular}{|p{0.28cm}|p{2.2cm}|p{2.2cm}|p{2.4cm}|}
\hline
& \textbf{original value} & \textbf{corrupted value} & \textbf{error magnitude} \\ \hline \hline
$itr_1$ & 0  &  0.000000059604645  &  $\infty$  \\ \hline
$itr_2$ & -0.004373951680278 & -0.004373951059397 & 6.20880999391282E-10  \\ \hline
$itr_3$  & -0.004816104396391 & -0.004816104262613 & 1.33777999962448E-10   \\ \hline
$itr_4$  & -0.004664456032917 & -0.004664455968072 & 6.48450000292899E-11  \\ \hline   
\end{tabular}
\end{center}
\squeezeup
\end{table}

We examine the value of the array element ($u[10][10][10]$) where a single bit-flip happens on the 40th bit
in the first invocation of the function $mg3P$. 
This function is iteratively called four times.
We examine error magnitude (as defined in 
Equation~\ref{eq:error}, recalling that error magnitude is the relative error of a faulty value). Table~\ref{tab:mg_repeatedly_adding} 
shows that the error magnitude becomes increasingly smaller  
as $mg3P$ is repeatedly called, reducing the effect of 
data corruption. Note that although the error magnitude at the 
second invocation of $mg3P$ is very small, it is still not acceptable 
for the verification phase of MG. However, as the corrupted value 
is closer to the correct value at the fourth invocation of $mg3P$, 
the corrupted value is acceptable by MG and regarded as a correct solution.


\begin{figure}[H]
\begin{lstlisting}[xleftmargin=.04\textwidth, xrightmargin=.01\textwidth]
/* find cluster center id with min dist to pt */
for (i=0; i<npts; i++) {
   float dist;
   dist = euclid_dist_2(pt, pts[i], nfeatures);  
   if (dist < min_dist) {
      min_dist = dist;
      index = i;
   }
}
\end{lstlisting}
\squeezeup
\caption{\small Example of the 
Conditional Statement pattern in KMEANS}
\label{fig:cond_statement}
\end{figure}

\begin{tcolorbox}[breakable,enhanced,left=1mm,right=1mm,top=0.75mm,bottom=0.75mm]
\small
\textbf{Pattern 3: 
Conditional Statements}\\
In this pattern, a conditional statement such as an \textit{if} condition, 
which tolerates a fault as long as the result of the statement in a faulty case remains the same (true/false) as in a fault-free case, consequently avoiding a control-flow divergence that otherwise could have occurred. 
The conditional statement can cause a decrease in the number of alive corrupted locations. 
\end{tcolorbox}

Although Pattern 3 is simple, it can become a major reason for 
fault tolerance in applications. KMEANS exemplifies this case:
Figure~\ref{fig:cond_statement} shows a code segment 
where a condition statement (Line 5) plays a major role to tolerate faults
in the array \textit{feature}. In essence, the code
tries to find the minimum distance between a target data point and 
the center data point of each cluster based on the feature values of data points. 
This conditional statement tolerates errors that happen in 
the array \textit{feature}, which takes most of the memory footprint of KMEANS. 
As long as the code segment can find the correct cluster with the minimum 
distance to the target point, the application outcome remains correct.

Besides the above example, we often find Pattern 3 in the program verification 
phases of MG and CG, where the final computation 
result is compared with a threshold to determine the result validity 
and/or to terminate execution. 




\begin{tcolorbox}[breakable,enhanced,left=1mm,right=1mm,top=0.75mm,bottom=0.75mm]
\small
\textbf{Pattern 4: Shifting} \\
In this pattern, bits are lost due to bit shifting operations.
If the lost bits are corrupted, fault tolerance 
occurs and we say that the pattern completely masks (or eliminates)
the faulty bit.
\end{tcolorbox}

We find Pattern 4 in IS---we show an example in 
Figure~\ref{fig:is_losing_data_bits}. IS is a benchmark that implements
bucket sorting for input integers (called ``keys'' in the benchmark).
The input integers are placed into multiple buckets based on their 
significant bits. To decide into which bucket a key will be placed, 
IS applies a shift operation on the key (Line 3 in 
Figure~\ref{fig:is_losing_data_bits}). If the data is corrupted in 
the least significant bits of the key, the shift operations can 
still correctly place the key into the corresponding bucket, 
hence tolerating faults in the key.

\begin{figure}
\begin{lstlisting}[xleftmargin=.04\textwidth, xrightmargin=.01\textwidth]
/*Determine the number of keys in each bucket*/
  for( i=0; i<NUM_KEYS; i++ )
    bucket_size[key_array[i] >> shift]++;
\end{lstlisting}
\squeezeup
\caption{\small Example of the shifting pattern in IS.}
\label{fig:is_losing_data_bits}
\end{figure}



\begin{tcolorbox}[breakable,enhanced,left=1mm,right=1mm,top=0.75mm,bottom=0.75mm]
\small
\textbf{Pattern 5: Data Truncation} \\
In this pattern, corrupted data is not presented to the 
user when used as a final result, or corrupted data is truncated.
	\end{tcolorbox}

We find Pattern 5 in LULESH, where in its last execution phase 
the computation results of a \textit{double} data type are reported 
in ``\%12.6e'' format (using the \texttt{printf} C function).
In this format, the mantissa of the computation 
result is partially cut-off and not fully presented to the user; 
thus if the cut-off mantissa is corrupted by a fault, the erroneous value
will not be seen by the user.

\begin{tcolorbox}[breakable,enhanced,left=1mm,right=1mm,top=0.75mm,bottom=0.75mm]
\small
\textbf{Pattern 6: Data Overwriting} \\
In this pattern, corrupted data is overwritten by a correct value, and the data corruption is consequently eliminated.
\end{tcolorbox}

We find Pattern 6 in all benchmarks, as it is commonly 
found in the output of many instructions. This occurs in particular
when the value of a corrupted location is overwritten by an
instruction that generates a clean uncorrupted value.

\subheader{Discussion.}
The effectiveness of some patterns (repeated additions, conditional statement, shifting, and data truncation) depends 
on the program input. For example, the effectiveness of 
the shifting pattern is dependent on the number of shifted 
bits---the more bits are shifted, the more 
random bit-flip errors can be tolerated. This is different 
from software design patterns that are general 
and independent of program input.

\section{Case Studies}
\label{sec:case_study}

Resilience computation patterns have 
many potential uses. We give two use cases. 
Here, whenever we use fault injection, 
we use 99\% confidence level and 1\% margin 
of error to decide the number of fault injection 
tests based on~\cite{date09:leveugle}.

\begin{table}[H]
\centering
\caption{Results after applying resilience patterns to CG.}
\label{tab:pattern_applied}
\scriptsize
\begin{tabular}{|p{2.1cm}|p{1.3cm}|p{3.2cm}|}
\hline
Resi. Pattern Applied     & App. Resi. & Exe time (s)/Average (s) \\ \hline
 None                & 0.59     & 158.659-159.468 / 159.010        \\ \hline
 DCL and overwrt.       & 0.78         & 158.859-159.457 / 159.167        \\ \hline
Truncation & 0.614        & 158.605-159.338 / 158.835        \\ \hline 
All together                      & 0.782        & 158.574-159.457 / 158.859        \\ \hline
\end{tabular}
\squeezeup
\end{table}

\subsection{Use Case 1: Resilience-Aware Application Design}
We apply resilience patterns to the CG benchmark, 
aiming to improve its resilience. We successfully apply three patterns: 
\textbf{dead corrupted location} (DCL), \textbf{data overwriting}, and 
\textbf{truncation}.
The results are shown in Table~\ref{tab:pattern_applied},
where the first column shows the resilience pattern(s) applied; the second column 
is the application resilience---the success rate measured by doing 
fault injection; the third column is the execution time for 
one run with or without applying resilience pattern(s). We report 
the average execution time for 20 runs in Table~\ref{tab:pattern_applied}.
Figures 12 and 13 in Appendix A show the 
code where we apply the three patterns.

To apply DCL and data overwriting, we introduce two temporal arrays 
at the beginning of $sprnvc()$ to replace two global arrays $v[]$ and $iv[]$ 
referenced in $sprnvc()$ (see Figure 12). Furthermore, to ensure the program correctness, 
the updated values of the two temporal arrays are copied back to $v[]$ and $iv[]$ 
at the end of $sprnvc()$.
Because of the copy-back, errors occurring 
in $v[]$ and $iv[]$ during the execution of $sprnvc()$ can be overwritten.
Moreover, errors that might occur in the two temporal 
arrays become dead (not accumulated as in the global arrays), after the copy-back.
Overall, we improve application resilience by 32.2\% with less than 
0.1\% performance loss (caused by a small amount of data movement).

To apply the truncation pattern, we select 10 iterations (340-350th iterations) 
of a loop within the function $conj\_grad()$, which is used to 
calculate $p\cdot q$ (see Figure 13). 
We replace 64-bit floating-point multiplications with 32-bit integer 
multiplications (particularly lines 508-510 in the source code). 
After applying the pattern, the precision loss (64 bit vs. 32 bit) 
does not affect the correctness of the final output. The reason is as follows.
As an iterative solver, CG gradually averages out the precision 
loss across iterations. Furthermore,
CG uses a conditional statement that compares the CG 
output with a threshold to verify the output correctness. 
Such conditional statement can further tolerate the precision loss.
Table~\ref{tab:pattern_applied} 
shows that we improve application resilience by 4.1\% with no performance loss. 
We apply the three patterns together and improve the application resilience 
by a total of \textbf{32.5\%} with less than 0.1\% performance loss.

\subsection{Use Case 2: Predicting Application Resilience}
The current common practice to quantify the resilience of an application
is to use random fault injection. 
However, random fault injection misses the application context that can explain 
how errors propagate and consequently are tolerated. 
In this case study, we are exploring a way alternative to random fault injection
to quantify application resilience. 
Since resilience computation patterns explain application
resilience, we may estimate the resilience of an application by counting the number of instances of such patterns
in the application. This approach can quantify the 
contribution of each resilience pattern to application resilience, which demonstrates the effectiveness of 
resilience patterns.

{\textbf{Model Construction.} 
We build a Bayesian multivariate linear regression model~\cite{BLR:Minka} to predict
the 
resilience (i.e., success rate) 
of an application.
The model uses the number of pattern instances for each resilience computation pattern 
as input, and outputs a single value $P_{suc\_rate}$, the predicted success rate. We 
model the above idea as follows:

\begin{equation}
\small
\label{eq:model}
P_{suc\_rate} = \sum_{i=1}^{\#patterns}\beta_{i}x_{i}+\epsilon.
\end{equation}
In Equation~\ref{eq:model}, $x_i$ is the number of pattern instances for a 
specific pattern $i$ normalized by total number of instructions within 
the application. We name $x_i$ the pattern rate (e.g., condition rate, shift rate, and truncation rate).
We normalize the number of pattern instances to enable a 
fair comparison between applications with different number of instructions. 
In total, there are $\#patterns$ patterns ($\#patterns$ is six in our modeling).
$\beta_i$ is the model coefficients and $\epsilon$ is the intercept.

\textbf{Experiments and Model Validation.}
We perform 
two experiments. In the \textit{first} experiment, we build 
the model using all the patterns from the ten benchmark programs (Section~\ref{sec:workloads}) to show
that the data fits the model well.
This experiment requires running the ten benchmarks, 
collecting the number of pattern instances for each pattern, and performing 
random fault injection to obtain success rates for each benchmark.

In the \textit{second} experiment, 
we train the model using data from different combinations of nine of the ten benchmarks, and make a prediction for success rate for the one remaining benchmark.
We then validate the model prediction by measuring its accuracy (i.e., relative error) with respect to the success rate that is obtained by doing fault injection. This experiment is to see how accurate the model is in predicting the success rate of an unseen program.

\begin{table*}[]
\centering
\caption{The quantification of resilience patterns and the prediction accuracy. SR=success rate}
\label{tab:prediction_error}
\scriptsize
\begin{tabular}{|p{1.4cm}|p{0.9cm}|p{1cm}|p{1cm}|p{1.2cm}|p{1.3cm}|p{1cm}|p{1cm}|p{1.5cm}|p{1cm}|}
\hline
  Benchmark   & Condition Rate & Shift Rate & Truncation Rate & Dead Location Rate & Repeat Addition Rate & Overwrite Rate & Measured SR & Predicted SR & Prediction Err. Rate \\
\hline
\hline
CG          & 0.088          & 2.45E-08   & 2.185           & 0.298              & 2.61E-07        & 0.999          & 0.739        &   0.652 & 11.8\%                     \\
\cline{1-10}
 MG          & 0.037          & 2.74E-03   & 1.145           & 0.314              & 0.000           & 0.999          & 0.879        &  0.810 & 7.8\%                     \\
\cline{1-10}
  LU          & 0.022          & 8.11E-06   & 0.188           & 0.319              & 0.000           & 0.999          & 0.575        &   0.642 & 11.7\%                    \\
\cline{1-10}
BT          & 0.015          & 0.000      & 0.074           & 0.334              & 0.000           & 0.999          & 0.656        &  0.573 & 12.7\%                     \\
\cline{1-10}
IS          & 0.040          & 2.86E-02   & 0.001           & 0.311              & 0.000           & 0.985          & 0.653        &  0.712 &  9.0\%                     \\
\cline{1-10}
DC          & 0.139 &	0.174 &	0.078 &	0.302 &	9.22E-07	& 0.994 &	0.578 &  0.204 & 64.6\%
\\
\cline{1-10}
SP          & 0.042          & 0.000      & 0.428           & 0.389              & 4.15E-08        & 0.999          & 0.385        &  0.466 & 21.0\%                     \\
\cline{1-10}
FT          & 0.038          & 1.99E-03   & 1.591           & 0.338              & 0.000           & 0.999          & 0.876        &  1.000 &  14.2\%                     \\
\cline{1-10}
KMEANS      & 0.079          & 7.18E-07   & 2.484           & 0.375              & 7.87E-05        & 0.979          & 0.843        &  1.000 & 18.6\%                     \\
\cline{1-10}
LULESH      & 0.048          & 2.60E-03   & 0.550           & 0.378              & 6.88E-06        & 0.937          & 0.926        &  0.725 & 21.7\%                     \\
\hline
\end{tabular}
\squeezeup
\end{table*}

\textbf{Experimental Results.}
For the first experiment, 
we calculate the ``$R-square$'' value of the model. 
$R-square$ is used for measuring the fitness of a statistic model.
The $R-square$ value in our experiment is 96.4\%, which is close to 1. A value close to 1 indicates that the model explains the variability of the prediction result around its mean. The model therefore fits and explains the data very well.

{For the second experiment, the prediction results 
are shown as the prediction error rate in Table~\ref{tab:prediction_error}.
The average prediction error excluding the prediction error 
on DC is 14.3\%. The prediction error on DC is large (64.6\%), because the model 
does not distinguish error tolerance capabilities of different instances 
of repeated additions and conditional statement (see the limitation discussed below), 
thus predictions for DC are affected by this limitation.

\textbf{Importance of Resilience Patterns: Feature Analysis.}
We use
standardized regression coefficient~\cite{SRC:Bring}, 
an indicator that presents the 
importance of predictors, to understand 
which resilience patterns are the most important. We compute
the standardized regression coefficients 
for the model trained in the second experiment.

On average, the \textit{averaged standardized regression 
coefficients} of Shifting, Truncation, Dead Location, Repeated Addition, Overwriting, 
and Conditional Statement are 1.48, 1.73, 0.38, 0.25, 0.92, and 1.69, respectively. 
\textit{We conclude that Truncation (1.73), Shifting (1.48), 
and Conditional Statement (1.69), that have the largest coefficients}, 
contribute the most to resilience.
On the other hand, patterns such as Repeated Addition 
and Dead Location have less impact.

\textbf{Limitation and Future Work.}
Different instances of a pattern can have different 
weight into application resilience.
For example, considering different cases of shifting where the value 
is shifted to right/left $x$ times. Depending on the value of $x$, 
the error may or may not be masked.
While simply counting the number of pattern instances
limits the prediction accuracy (one should also take into account the  
value of locations), this demonstrates 
a simple but practical use case of the patterns.

\section{Related Work}
\label{sec:related_work}
\subheader{Resilience Computation Patterns.}
A limited number of previous studies reveal the existence of resilience 
patterns~\cite{sc16:guanpeng, dsn08:cook}; these efforts, however, lack a systematic
method to identify these patterns. In~\cite{sc16:guanpeng}, Li et al. 
identify conditional statement and truncation for error masking 
in GPU programs. In~\cite{dsn08:cook}, Cook and Zilles identify shift, 
conditional statement and truncation. Those research efforts
manually examine fault tolerance cases, while
our work is different in several aspects.
First, we introduce a novel \textit{framework and methodology 
to systematically} identify patterns.
For complex applications, manual identification of those patterns is 
unfeasible. Second, we identify more complex patterns (e.g., 
DCL and repeated additions). 
Those new patterns require multiple instructions to take effect.
Finding those patterns must be based on a complete picture on 
error propagation. The existing work identifies patterns based 
on the analysis of individual instructions without sufficient 
considerations of interactions between instructions, 
hence lacking a complete picture to identify patterns. 

\subheader{Error Detector Placement.}
Existing research uses compiler static and/or 
dynamic instruction analysis to enable application-level fault tolerance 
by detecting code vulnerabilities. For example, Pattabiraman et al. 
use static analysis~\cite{tdsc11:pattabiraman} and a data-dependence 
analysis~\cite{prdc05:karthik} to determine the placement of error 
detectors in applications. Their work determines the critical variables 
that are likely to propagate errors based on metrics, such as  highest 
dynamic fan-out. Different from us, their work cannot locate resilience patterns.

\subheader{Visualization.}
Recently, techniques that allow visualization of corrupted application
data across loop iterations and MPI processes have been developed.
For example, Calhoun et al.~\cite{hpdc17:Calhoun} replicate instructions
to track and visualize how errors propagate within the application.
However, their approach can be expensive when analyzing complex applications. Our approach, based on the abstract code structure model, can accelerate tracking error propagation.
\section{Conclusions}
Understanding natural error resilience in HPC applications
is important in creating applications that can naturally
tolerate errors. 
%
However, our knowledge on natural error resilience has been 
quite limited, mainly because of a lack of systematic
methods to identify resilience computation patterns.
Our framework, \name, exposes these patterns by
enabling fine-grained tracking of 
error propagation and fault tolerance to enable users
to pinpoint resilience computations in HPC programs.
By tracking data flows and value variations 
based on a code region model, we identify
and summarize six common resilience patterns, which 
increase our understanding of how natural resilience occurs.
We also present two case studies of practical
applications of these resilience patterns.

\section*{Acknowledgements}
We thank anonymous reviewers for their valuable feedback.
This work was performed under the auspices of the U.S. Department of
Energy by Lawrence Livermore National Laboratory under contract
DEAC52-07NA27344 (LLNL-CONF-748619).
This work is partially supported by U.S. 
National Science Foundation (CNS-1617967, CCF-1553645 and CCF-1718194).

\bibliographystyle{IEEEtran}
\bibliography{az}

\clearpage
\newpage\section{Appendix}
\label{sec:app}

\subsection{\textbf{Details of Use Case 1: Resilience-Aware Application Design}}
\label{app:case_1}
Figure~\ref{fig:case1_dcl} and Figure~\ref{fig:case1_trunc} show 
two code excerpts extracted from CG, where dead corrupted location, data 
overwriting and truncation are applied, respectively.

For the case of dead corrupted location and data overwriting,
the original code is shown in Figure~\ref{fig:case1_dcl}(a)
and the new code is shown in Figure~\ref{fig:case1_dcl}(b)
(we include some comments to explain the difference). In particular, we use two 
temporal arrays $v\_tmp$ and $iv\_tmp$ to replace two global arrays 
$v$ and $iv$. We then copy values in the arrays $v\_tmp$ and $iv\_tmp$ 
back to the arrays $v$ and $iv$ after the computation. 

\begin{figure*}[!hbp]
\begin{minipage}[]{0.46\linewidth}
\centering
\lstset{language=C}
\begin{lstlisting}[xleftmargin=.06\textwidth, frame=single, numbers=left, mathescape, title={(a)}]
static void sprnvc(int n, int nz, int nn1, double v[], int iv[]){
  int nzv, ii, i;
  double vecelt, vecloc;
  
  
  
  
  
  
  nzv = 0;
  while (nzv < nz) {
    vecelt = randlc(&tran, amult);
    vecloc = randlc(&tran, amult);
    i = icnvrt(vecloc, nn1) + 1; 
    if (i > n) continue;
    logical was_gen = false;
    for (ii = 0; ii < nzv; ii++) {
      if (iv[ii] == i) { 
        was_gen = true;
        break;
      }
    }
    if (was_gen) continue;
    v[nzv] = vecelt; 
    iv[nzv] = i; 
    nzv = nzv + 1;
  }
  
  
  
  
}
\end{lstlisting}
\end{minipage}
\hspace{0.5cm}
\begin{minipage}[]{0.46\linewidth}
\centering
\begin{lstlisting}[frame=single, numbers=left, mathescape,%
   title={(b)}]
static void sprnvc(int n, int nz, int nn1, double v[], int iv[]){
  int nzv, ii, i;
  double vecelt, vecloc;
  double v_tmp[NONZER+1]; //define a temp array
  int iv_tmp[NONZER+1];  //define a temp array
  for(i=0;i<=NONZER;i++){ 
    v_tmp[i] = v[i];  //initialization      
    iv_tmp[i] = iv[i]; //initialization 
  } 
  nzv = 0;
  while (nzv < nz) {
    vecelt = randlc(&tran, amult);
    vecloc = randlc(&tran, amult);
    i = icnvrt(vecloc, nn1) + 1; 
    if (i > n) continue;
    logical was_gen = false;
    for (ii = 0; ii < nzv; ii++) {
      if (iv_tmp[ii] == i) { //replace iv with iv_tmp
        was_gen = true;
        break;
      }
    }
    if (was_gen) continue;
    v_tmp[nzv] = vecelt;  //replace v with v_tmp
    iv_tmp[nzv] = i;  //replace iv with iv_tmp
    nzv = nzv + 1;
  }
  for(i=0;i<=NONZER;i++){ 
    v[i] = v_tmp[i];  //copy back
    iv[i] = iv_tmp[i]; //copy back
  } 
}
\end{lstlisting}
\end{minipage}
\caption{\small A code excerpt from the function $sprnvc$() in CG for the Use Case 1. (a) shows the original code excerpt before patterns are applied; (b) shows the code excerpt when dead corrupted location and data overwriting are applied. }
 \label{fig:case1_dcl}
 \squeezeup
\end{figure*}

\begin{figure*}[h]
\begin{minipage}[]{0.46\linewidth}
\centering
\lstset{language=C}
\begin{lstlisting}[xleftmargin=.08\textwidth, frame=single, numbers=left, mathescape, title={(a)}]
static void conj_grad(int colidx[],
					  ...
                      double p[],
                      double q[])
{
  ...
  // Obtain p.q 
  d = 0.0;
  for (j = 0; j < lastcol - firstcol + 1; j++) {
  
  
  
  
  
    d = d + p[j]*q[j];
    
  }
  ...
}
\end{lstlisting}
\end{minipage}
\hspace{0.5cm}
\begin{minipage}[]{0.46\linewidth}
\centering
\begin{lstlisting}[frame=single, numbers=left, mathescape,%
   title={(b)}]
static void conj_grad(int colidx[],
					  ...
                      double p[],
                      double q[])
{
  ...
  // Obtain p.q 
  d = 0.0;
  for (j = 0; j < lastcol - firstcol + 1; j++) {
    if(j<=350&&j>=340){
      int tmp = p[j]; // truncation
      int tmp1 = q[j]; // truncation
      d = d + tmp*tmp1;
    }else{
      d = d + p[j]*q[j];
    }
  }
  ...
}
\end{lstlisting}
\end{minipage}
\caption{\small A code excerpt from the function $conj\_grad()$ in CG for the Use Case 1. (a) shows the original code excerpt before the truncation pattern is applied; (b) shows the code excerpt when the truncation is applied.}
 \label{fig:case1_trunc}
 \squeezeup
\end{figure*}

Figure~\ref{fig:case1_trunc} shows how we apply the truncation. 
In particular, we replace 64-bit floating-point multiplications to 32-bit integer multiplications (see Lines 11-12 in    
Figure~\ref{fig:case1_trunc}.b).

\end{document}